\documentclass[9pt,twocolumn,twoside]{pnas-new} % ,lineno
\templatetype{pnasinvited} % Choose template 
\setboolean{displaywatermark}{false}
\usepackage{hyperref}
\usepackage{lettrine}

\title{DenseUNets with feedback non-local attention for the segmentation of specular microscopy images of the corneal endothelium with guttae}

\author[a]{Juan P. Vigueras-Guill\'{e}n}
\author[b]{Jeroen van Rooij} 
\author[c,d]{Bart T.H. van Dooren}
\author[b]{Hans G. Lemij}
\author[b]{Esma Islamaj}
\author[a]{Lucas J. van Vliet}
\author[e]{Koenraad A. Vermeer}

\affil[a]{Department of Imaging Physics, Delft University of Technology, Delft, The Netherlands}
\affil[b]{Rotterdam Eye Hospital, Rotterdam, The Netherlands}
\affil[c]{Amphia Hospital, Breda, The Netherlands}
\affil[d]{Erasmus Medical Center Rotterdam, Rotterdam, The Netherlands}
\affil[e]{Novo Research Consultancy, Voorburg, The Netherlands}

\leadauthor{Vigueras-Guill\'{e}n} 

\authorcontributions{This work was supported by the Dutch Organization for Health Research and Healthcare Innovation (ZonMw, The Hague, The Netherlands) under Grants 842005004 and 842005007, and by the Combined Ophthalmic Research Rotterdam (CORR, Rotterdam, The Netherlands) under grant no. 2.1.0.}
\authordeclaration{The authors declare that they have no competing interests.}
\equalauthors{} %\textsuperscript{1}A.O.(Author One) contributed equally to this work with A.T. (Author Two) (remove if not applicable).
\correspondingauthor{Juan P. Vigueras-Guill\'{e}n. E-mail: J.P.ViguerasGuillen@gmail.com} 


\begin{abstract}
To estimate the corneal endothelial parameters from specular microscopy images depicting cornea guttata (Fuchs dystrophy), we propose a new deep learning methodology that includes a novel attention mechanism named feedback non-local attention (fNLA). Our approach first infers the cell edges, then selects the cells that are well detected, and finally applies a postprocessing method to correct mistakes and provide the binary segmentation from which the corneal parameters are estimated (cell density [ECD], coefficient of variation [CV], and hexagonality [HEX]). In this study, we analyzed 1203 images acquired with a Topcon SP-1P microscope, 500 of which contained guttae. Manual segmentation was performed in all images. We compared the results of different networks (UNet, ResUNeXt, DenseUNets, UNet++) and found that DenseUNets with fNLA provided the best performance, with a mean absolute error of 23.16 [cells/mm$^{2}$] in ECD, 1.28 [\%] in CV, and 3.13 [\%] in HEX, which was 3--6 times smaller than the error obtained by Topcon’s built-in software. Our approach handled the cells affected by guttae remarkably well, detecting cell edges occluded by small guttae while discarding areas covered by large guttae. Overall, the proposed method obtained accurate estimations in extremely challenging specular images. 
\end{abstract} 

\dates{} %This manuscript was compiled on \today
\doi{} %\url{www.pnas.org/cgi/doi/10.1073/pnas.XXXXXXXXXX}

\begin{document}

\maketitle
\thispagestyle{firststyle}
\ifthenelse{\boolean{shortarticle}}{\ifthenelse{\boolean{singlecolumn}}{\abscontentformatted}{\abscontent}}{}

%\section{Introduction}
% \noindent 
\lettrine[lines=2]{F}{uchs} endothelial dystrophy (FED) is a corneal disease characterized by an increase in thickness of the Descemet's membrane, a large deposition of extracellular matrix in the corneal endothelium (referred as guttae), and a progressive loss of endothelial cells \cite{Elhalis2010}. The guttae, which are condensation of collagen growing from the Descemet's membrane (\textit{gutta} is the Latin word for `droplet', being \textit{guttae} the plural form and \textit{guttata} the adjective form), appear disperse in the early stages of the disease, but as FED progresses, guttae become abundant and visible in the entire endothelium \cite{McLaren2014}. This is accompanied by a great loss of endothelial cells that is irreversible due to the lack of cell regeneration. Eventually, the endothelium is not able to support corneal deturgescence, thereby leading to corneal edema, poor visual acuity, and finally the necessity for corneal transplantation to restore vision. FED is the most common cause for the transplantation of the corneal endothelium worldwide \cite{Gain2016}. This disease usually appears in people of 40--60 years of age and it has a slow clinical progression (10--20 years) \cite{Adamis1993}. Since many patients with cornea guttata do not progress to FED, it is difficult to assess its incidence and prevalence. Nevertheless, it is estimated that around 4\% of the population over 40 years of age have corneal guttae \cite{Moshirfar2021}. 

The most common staging system used to describe FED has four levels \cite{Elhalis2010,Adamis1993}: in stage 1, guttae appear in the central cornea and they are non-confluent; in stage 2, guttae start to coalesce and spread towards the peripheral cornea, while endothelial cell loss accelerates; in stage 3, stromal edema is present, which might cause the formation of epithelial and subepithelial bullae; and in stage 4, the cornea has become opaque due to chronic edema and visual acuity is critically compromised. Conversely, central guttae can be observed in the elderly without the presence of edema, and the existence of only peripheral guttae is related to a condition called `Hassall-Henle bodies' and does not lead to edema \cite{Elhalis2010}. An endothelial cell density (ECD) between 400 to 700 cells/mm$^{2}$ tends to result in corneal decompensation \cite{Foster2004}. Thus, the correct estimation of ECD, along with the other two main endothelial parameters (polymegethism or coefficient of variation in cell size [CV], and pleomorphism or hexagonality [HEX]), is a valuable tool for the diagnosis and monitoring of this dystrophy.   % (this is usually not classified as FED)

\begin{figure}[t!] %[tbhp]
	\centering
	\includegraphics[width=1\linewidth]{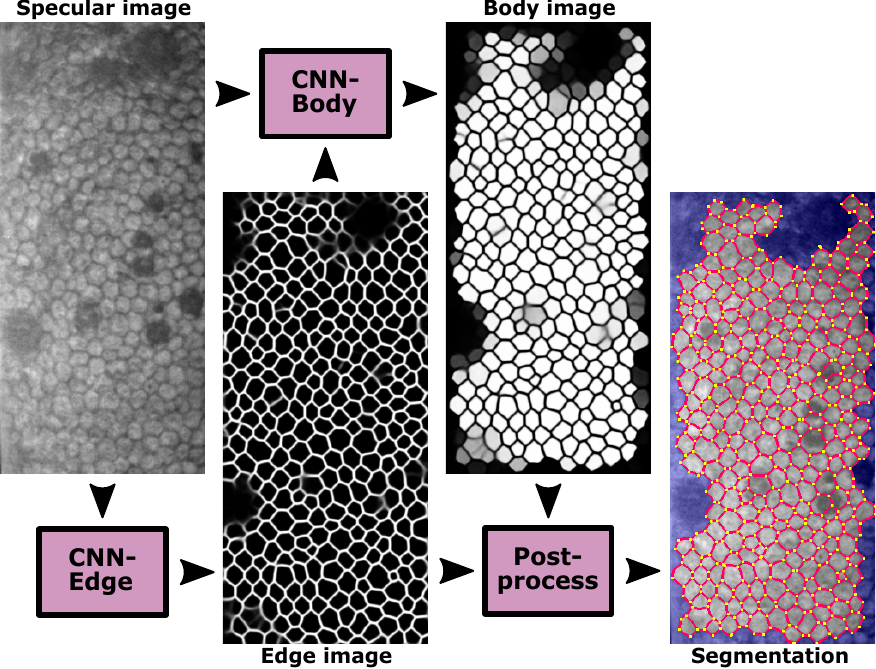}
	\caption{Flowchart: CNN-Edge detects the cell edges from the specular image, CNN-Body infers the cell bodies that are well detected, and the postprocessing refines the edge images and applies the ROI from the body images to provide the final segmentation. The final segmentation (edges in red, vertices in yellow, non-ROI area in blue) is dilated and superimposed onto the specular image for illustrative purposes.}
	\label{fig01}
\end{figure}

The endothelium can be easily imaged with specular microscopy, a noninvasive, noncontact method that relies on the reflection of light from the interface between the endothelium and aqueous humor \cite{McCarey2008}. However, it is crucial to have a smooth, regular endothelial surface to provide good-quality images. In this respect, the guttae push the endothelial cells out of the specular plane and, thus, these appear as `black droplets' in the specular images (Fig.~\ref{fig01}). If the guttae are non-confluent and have a relatively small size (similar or less than the average cell size), a human observer could probably infer the cell tessellation in the area. However, manual annotations are very tedious and time-consuming, and there are currently no automatic methods that can perform such inference. In fact, many recent studies have shown that today's commercial methods provided by the microscope manufacturers to automatically segment the endothelial images depict low reliability, even for healthy corneas \cite{Huang2018, Price2013, Luft2015, Gasser2015, Kitzmann2004}. 

In recent years, several new approaches to estimate the corneal endothelial parameters have been proposed. Up to 2018, the methods were based on image processing techniques and classic machine learning \cite{Piorkowski2015, Selig2015, Scarpa2016, Al-Fahdawi2018, Vigueras2018}, but from 2018 onward, a significantly large number of new approaches based on deep learning have been presented \cite{Fabijanska2018, Nurzynska2018, Daniel2019, Fabijanska2019, Kolluru2019, Vigueras2019a, Vigueras2019b, Vigueras2019c, Joseph2020, Sierra2020, Vigueras2020, Karmakar2021, Kucharski2021, Shilpashree2021, Herrera2021}. Overall, these methods have shown a good performance. However, it is worth noting a few details: (i) some methods (mainly from the pre-deep learning era) evaluated the estimation of the three corneal parameters but the images were from healthy eyes and there was either a manual selection of the region of interest (ROI) or the cells were visible in the whole image \cite{Selig2015, Scarpa2016, Al-Fahdawi2018, Vigueras2018, Fabijanska2018, Fabijanska2019, Vigueras2019a}; (ii) some of the first deep learning methods simply evaluated the capacity of neural networks to perform an accurate segmentation, either in healthy \cite{Nurzynska2018, Kolluru2019} or unhealthy corneas \cite{Joseph2020}, without estimating any corneal parameter, which avoids the non-trivial problem of refining the raw segmentation for the purpose of obtaining an accurate biomarker estimation; (iii) other publications only focused in estimating ECD (or the number of cells) in healthy cases \cite{Karmakar2021, Kucharski2021} and images with guttae \cite{Daniel2019, Sierra2020, Shilpashree2021}, some including a method to select the ROI (although this part is often unclear); and (iv) our previous work is, to the best of our knowledge, the only fully-automatic method to estimate all three parameters in all types of images (heavily blurred \cite{Vigueras2019b, Vigueras2019c} and also with some guttae \cite{Vigueras2020}). Among the publications dealing with guttae, a quick visual inspection is enough to perceive the inaccurate segmentation around the guttae, where partial cells occluded by the guttae are included as full cells \cite{Sierra2020, Shilpashree2021}; in contrast, our previous work \cite{Vigueras2020} has shown better results but still failed in cases of very advanced disease. Therefore, the accurate segmentation of images in the presence of guttae is still an unsolved problem. 

In this study, we present a new approach that automatically segments endothelial images with guttae and estimates the endothelial parameters. The proposed method contains three subprocesses (Fig.~\ref{fig01}): (i) a convolutional neural network (CNN) that infers the cell edges in the image (named CNN-Edge, yielding an `edge image'); (ii) another CNN model that infers the cells that can be fully identified (named CNN-Body, yielding a [cell] `body image'); and (iii) an image processing model based on watershed that refines the edge images, uses the body images to select the well-detected cells, and extracts the corneal parameters (named postprocessing). A comparative analysis is performed to determine what type of deep learning model (UNet \cite{Ronneberger2015}, ResUNeXt \cite{Xie2017}, DenseUNet \cite{Jegou2017}, and UNet+/++ \cite{Zhou2020}) is more appropriate for this task. We also present a new attention mechanism, named feedback non-local attention (fNLA), which can be plugged to any of the aforementioned networks. Different versions of the attention mechanism are tested, and we compare it to a well-know `attention UNet' \cite{Oktay2018}. This framework is evaluated against our previous approach \cite{Vigueras2020}, named CNN-ROI, which is retrained with the dataset of this study. A variation of CNN-Body, named CNN-Blob, which infers the overall area of interest instead of independent cells, is also evaluated. Finally, these frameworks are compared against the manual annotations and the estimates provided by the microscope's software.

\section*{Materials \& Methods}

\subsection*{Datasets}

Two datasets were employed in this work. The first dataset came from a clinical study concerning the implantation of a Baerveldt glaucoma implant in the Rotterdam Eye Hospital (Rotterdam, the Netherlands), with trial registration NL4823 (\url{https://www.trialregister.nl/trial/4823}, registered on 06-01-2015). The clinical study contained 7975 images from 204 patients (average age 66$\pm$10 years), who were imaged with a specular microscope (Topcon SP-1P, Topcon Co., Tokyo, Japan) before the device implantation and at 3, 6, 12, and 24 months after, in both the central (CE) and the peripheral supratemporal (ST) cornea. The protocol required five specular images to be taken in each area for each visit, although it was sometimes difficult to reach that number of gradable images (specifically, an average of 4.7 CE images and 3.6 ST images per visit were acquired). Written informed consent was obtained from all participants, and the study was conducted in accordance with the principles of the Declaration of Helsinki (October, 2013). Retrospectively, we observed that 15 patients had clear signs of FED stage two: they all had guttae in both CE and ST cornea, with a larger amount in CE and a clear increase in ST during the two-year follow-up. From this subset of patients, 193 images were collected. Furthermore, we observed that another 81 patients had small amounts of non-confluent guttae in the CE cornea (either FED stage one or due to normal aging), and 227 images were collected from these cases. In total, 420 images with presence of guttae were selected from this study. In addition, 400 images from other patients in the study without signs of guttae were selected to build a balanced dataset. 

The second dataset came from another clinical study in the Rotterdam Eye Hospital regarding the transplantation of the cornea (ultrathin Descemet Stripping Automated Endothelial Keratoplasty, UT-DSAEK), with trial registration NL4805 (\url{https://www.trialregister.nl/trial/4805}, registered on 15-12-2014). This dataset contained 383 images of the central cornea from 41 eyes (41 patients, average age 73$\pm$7 years) and they were acquired at 1, 3, 6, and 12 months after surgery with the same specular microscope Topcon SP-1P. The included population for the study were patients over 18 years old with FED planned for keratoplasty. Written informed consent was obtained from all participants. Among these patients, FED reappeared in one of them. Another 13 patients showed a small amount of non-confluent, stable guttae during the one-year follow-up. In total, 80 images out of the 383 showed some guttae (all images were included in the present work).

Altogether, the combined dataset contained 1203 images, in which 500 images depicted guttae in various magnitudes. The images covered an area of approximately 0.25 mm × 0.55 mm and were saved as 8-bit grayscale images of 240×528 pixels. All images were manually segmented to create the gold standard, using the open-source image manipulation program GIMP (version 2.10). Furthermore, we collected the endothelial parameters provided by the microscope (Topcon SP-1P performed this with the software IMAGEnet i-base, version 1.32).

\subsection*{Grading the dataset}

The 500 images with guttae were graded based on their complexity to segment them, taking two metrics into account: the amount of guttae and blur present in the image. For both metrics, images were given a value of 1 (mild), 2 (moderate), or 3 (severe), the final grade being the sum of both values. As a result, there were 134 images with low complexity (grades 1--2), 235 images with medium complexity (grades 3--4), and 131 images with high complexity (grades 5--6).

\begin{figure*}[!t]
	\centering
	\includegraphics[width=1\linewidth]{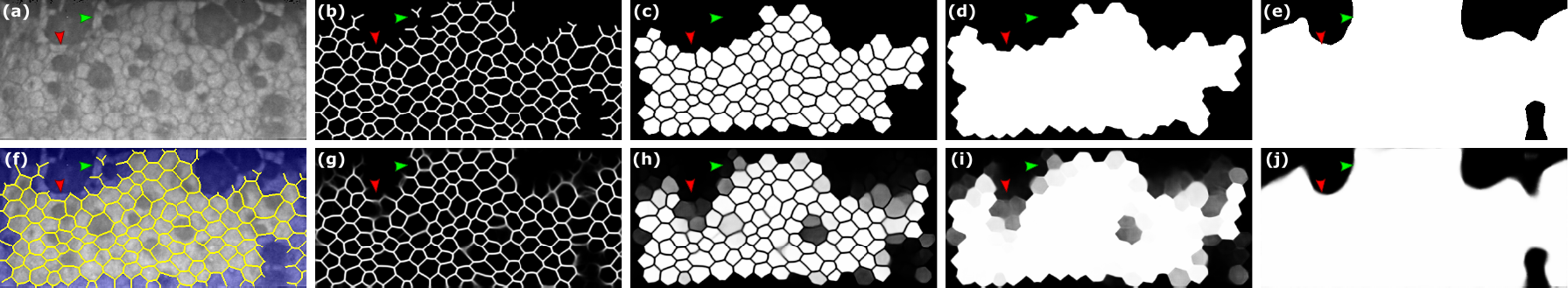}
	\caption{\textbf{(a)} Illustrative specular image with guttae (grade 4) and \textbf{(f)} its manual annotation (the edges, in yellow, are dilated for illustrative purposes; non-ROI area in blue). For this image, the targets in the different networks are: \textbf{(b)} CNN-Edge, \textbf{(c)} CNN-Body, \textbf{(d)} CNN-Blob, and \textbf{(e)} CNN-ROI. The outcome of the networks during inference (DenseUNet fNLA-Mul) are: \textbf{(g)} CNN-Edge, \textbf{(h)} CNN-Body, \textbf{(i)} CNN-Blob, and \textbf{(j)} CNN-ROI. Red and green arrows indicate two peculiarities of the targets (to be discussed later). }
	\label{fig02}
\end{figure*}

\subsection*{Targets and frameworks}
CNN-Edge is the core of the method. If the specular image had good quality (high contrast) and with cells visible in the whole image, the resulting edge image would probably be so well inferred that a simple thresholding and skeletonization would suffice to obtain the binary segmentation. However, these issues (low contrast, blurred areas, and guttae) are present in the current images. In contrast, CNN-Body has the goal of providing a ROI image to discard areas masked by extensive guttae or blurriness. 

These CNNs are trained independently and they all have the same design; thus, they are simply trained with different inputs and targets. To create the targets, we make use of the manual annotations (i.e. gold standard), which are binary images where value 1 indicates a cell edge (edges are 8-connected-pixel lines of 1 pixel width), value 0 represents a full cell body, and any area to discard (including partial cells) is given a value 0.5. If a blurred or guttae area is so small that the cell edges could be inferred by observing the surroundings, the edges are annotated instead (Fig.~\ref{fig02}-f). For all the annotated cells, we identify their vertices; this allows computing the parameter hexagonality from all cells and not only the inner cells (in the latter, HEX is computed by counting the neighboring cells, thus the cells in the periphery of the segmentation are not considered; this way of computing HEX was used in previous publications \cite{Vigueras2018b,Vigueras2019a,Vigueras2019b,Vigueras2019c,Vigueras2020} and it is how Topcon's built-in software computes it). Therefore, HEX is now defined as the percentage of cells that have six vertices.

The target of the CNN-Edge only contains the cell edges from the gold standard images, which have been convolved with a 7$\times$7 isotropic unnormalized Gaussian filter of standard deviation (SD) 1 (Fig.~\ref{fig02}-b). This provides a continuous target with thicker edges, which proved to deliver better results than binary targets \cite{Vigueras2019a}.

The target of the CNN-Body only contains the full cell bodies from the gold standard images, and partial cells are discarded either because they are partially occluded by large guttae or they are at the border of the image (Fig.~\ref{fig02}-c). The same probabilistic transformation is applied here. Alternatively, we also evaluated whether a target that also includes the edges between the cell bodies (Fig.~\ref{fig02}-d) was a better approach (named CNN-Blob). % Fig.~\ref{fig02} illustrates how the targets of CNN-Edge and CNN-Body are not complementary to each other. 

This framework is similar to our previous approach \cite{Vigueras2020}, where a model named CNN-ROI, whose input is the edge image, provides a binary map indicating the ROI (Fig.~\ref{fig02}-j). To create its target, the annotator would simply draw the rough area from the edge images (Fig.~\ref{fig02}-g) that they would choose as trustworthy, creating a binary target (Fig.~\ref{fig02}-e).

\subsection*{Design of the Convolutional Neural Networks}

\begin{figure*}[t]
	\centering
	\includegraphics[width=1\linewidth]{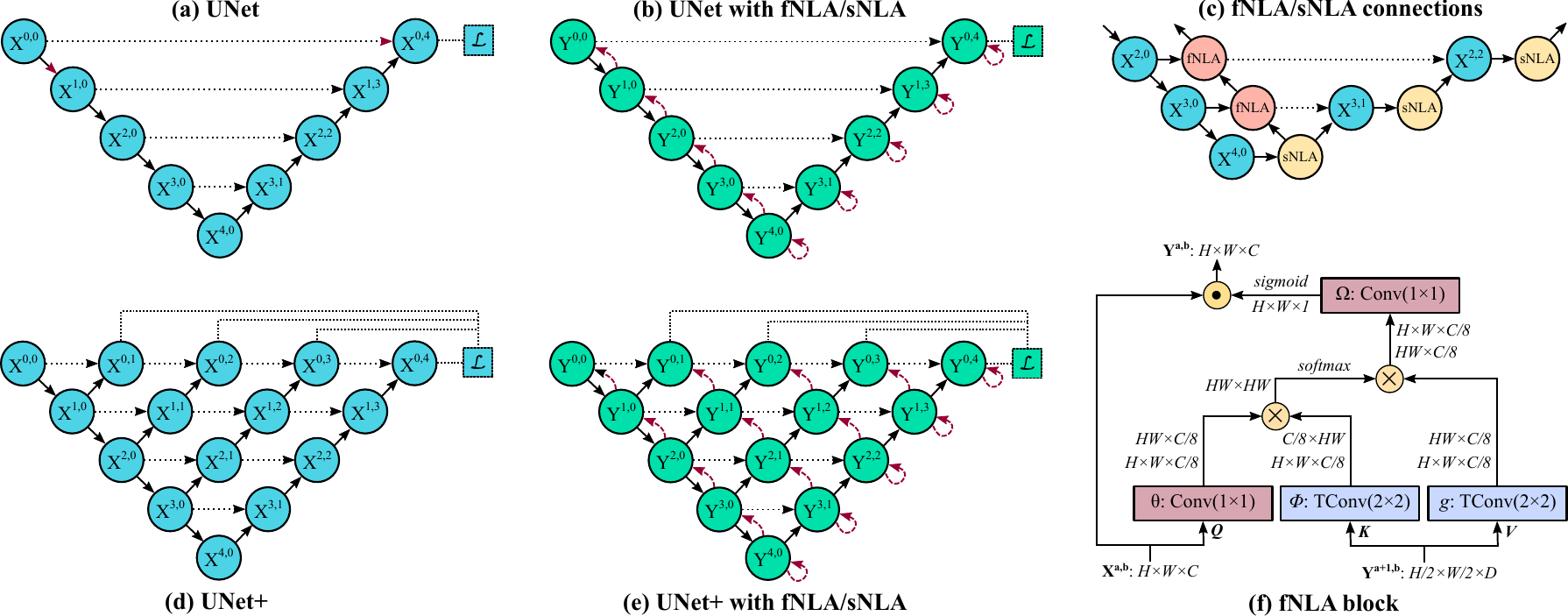}
	\caption{A schematic representation of the \textbf{(a)} UNet backbone and \textbf{(d)} its equivalent UNet+; a simplified representation of adding fNLA/sNLA blocks to the \textbf{(b)} UNet and \textbf{(e)} UNet+, where a red arrow indicates self-attention (sNLA) if it returns to the same node or feedback attention (fNLA) otherwise; \textbf{(c)} a schematic overview of how the fNLA/sNLA blocks are added to the UNet backbone for the three deepest resolution scales; and \textbf{(f)} a detailed description of an fNLA block with multiplicative aggregation: the feature maps are shown as the shape of their tensors, $X^{a,b}$ is the tensor to be transformed, $Y^{a+1,b}$ is the tensor from the lower resolution scale used for attention, the blue boxes ($\phi$ and $g$) indicate a 2×2 transpose convolution with strides 2, the red boxes ($\theta$ and $\Omega$) indicate a 1×1 convolution, $\oplus$ denotes matrix multiplication, and $\odot$ denotes element-wise multiplication. In the case of sNLA, $\phi$ and $g$ are also a 1×1 convolution with $X^{a,b}$ as input. In attention terminology \cite{Wang2018}, $Q$ is for query, $K$ for key, $V$ for value. }
	\label{fig03}
\end{figure*}

The backbone of the proposed CNN has five resolution stages (Fig.~\ref{fig03}-a). In this work, we tested three different basic designs depending on the connections within each node: the convolutional layers can be connected consecutively (as in UNet \cite{Ronneberger2015}), they may use residual connections (ResUNeXt \cite{Xie2017}) or dense connections (DenseUNet \cite{Jegou2017}). We also explored two multiscale designs, UNet+ (Fig.~\ref{fig03}-d) and UNet++ \cite{Zhou2020}, for the three types of networks. Our UNet++ differentiates from UNet+ in that the former uses feature addition from all previous short connections (transition blocks) of the same resolution stage. In total, nine basic networks were tested (the code for all networks can be found in our GitHub).

Here, we briefly describe the network that provided the best performance, the DenseUNet. In this network, the dense nodes have 4, 8, 12, 16, or 20 convolutional blocks, depending on the resolution stage, with a growth rate (GR) of 5 feature maps per convolutional layer. Each convolutional block has a compression layer, Conv(1×1, 4$\cdot$GR maps) + BRN\cite{Ioffe2016} + ELU\cite{Clevert2016}, and a growth layer, Conv(3×3, GR maps) + BRN + ELU, followed by Dropout(20\%) + Concatenation (with block's input), except the first dense block, which lacks the compression layer and dropout \cite{Srivastava2014} (all nodes in the first resolution stage lack dropout). The transition layer (short-connections) has Conv(1×1, $\alpha$ maps) + BRN + ELU, the downsampling layer has Conv(2×2, strides 2, $\alpha$ maps) + BRN + ELU, and the upsampling layer has ConvTranspose(2×2, strides 2, $\alpha$ maps) + BRN + ELU, being $\alpha$=(number of blocks in previous dense block)×GR/2. The output of the last dense block, $X^{04}$, uses a transition layer with 2 maps to provide the output of the network. 
% (BRN indicates batch renormalization and ELU is the type of activation)

Furthermore, we propose a new attention block to be added at the end of each dense block, named feedback non-local attention (fNLA, Fig.~\ref{fig03}-f) or self-non-local attention (sNLA), depending where it is placed within the network (Fig.~\ref{fig03}-c): if there exists a tensor from a lower resolution stage, the attention mechanism makes use of it (fNLA), but in the absence of such lower tensor, a self-attention operation is performed (sNLA, where $\phi$ and $g$ in Fig.~\ref{fig03}-f become a 1×1 convolution with the same input $X^{a,b}$). In the DenseUNet, the nodes of the encoder use fNLA (except the deepest node), and the nodes of the decoder use sNLA (Fig.~\ref{fig03}-b); in the DenseUNet+/++, only the largest decoder use sNLA (Fig.~\ref{fig03}-e). 

We explored three types of fNLAs, depending on the type of aggregation at the end of the block: (a) the default case (Fig.~\ref{fig03}-f) uses multiplicative aggregation, where a single attention map (from $\Omega$) is sigmoid activated and then element-wise multiplied to all input feature maps; (b) in case of additive aggregation, $\Omega$ becomes a 1×1 convolution with ReLU activation and C output feature maps, which are then summed to the input tensor; (c) in the case of concatenative aggregation, $\Omega$ is simply an ELU activation, whose output maps (C/8 maps) are concatenated to the input features. 

The design of this attention block is inspired by Wang et al.'s non-local attention method \cite{Wang2018}, which also resembles a scaled dot-product attention block \cite{Vaswani2017, Srinivas2021}. Basically, Wang et al.'s model \cite{Wang2018} is an sNLA with additive aggregation. Intuitively, an sNLA mechanism computes the response at a specific point in the feature maps as the weighted sum of the features at all positions. In fNLA, the attention block maps the input tensor against the output features from the lower dense block, thus allowing the attention mechanism to use information created further ahead in the network. Moreover, the feedback path created in the encoder allows to propagate the attention features from the deepest dense block back to the first block. While endothelial specular images do not possess such long-range dependencies (features separated by a distance of 3--4 cells do not seem to be correlated), this attention operation might be useful in the presence of large blurred areas (such as guttae). In this work, we explored the use of this attention mechanism for DenseUNet and the multiscale versions DenseUNet+/++. %, and we compared it with a well-known attention UNet proposed by Oktay et al. \cite{Oktay2018}.

\subsection*{Description of the postprocessing}

The postprocessing aims to fix any edge discontinuity. Here, we have improved a process that was first described in a previous publication \cite{Vigueras2019a}. The steps are:

\textbf{(I)} We estimate the average cell size ($l$) by using Fourier analysis in the edge image \cite{Vigueras2019a}. As we proved in previous work \cite{Vigueras2019c},  this estimation is simple, extremely robust, and accurate. 

\textbf{(II)} As new step, we add a perimeter to the edge image with intensity 0.5. This closes any partial cell in touch with the border. 

\textbf{(III)} We smooth the edge image with a Gaussian filter whose standard deviation is $\sigma=k_{\sigma}l$, being $k_{\sigma}=0.2$ (this parameter was derived in a previous publication \cite{Vigueras2018b}). This fixes any edge discontinuity. % Then

\textbf{(IV)} We apply classic watershed \cite{Beucher1993} to the smoothed edge image, which detects weak edges. This provides a binary segmentation where edges are 8-connected-pixel lines of 1 pixel width. % Afterwards

\textbf{(V)} In another new step, we identify every edge and vertex in the segmentation. The vertices are the branch points of the segmentation, and the edges are the set of 8-connected positive pixels whose endpoints are constrained to vertices \cite{Vigueras2018}. We set 2 pixels as the minimum length for an edge (edges of only 1 pixel are fused with the vertices at its endpoints to become a single vertex of 3 pixels). For every edge, we check its mean intensity (from the edge images) to discard the weak ones: if it is lower than 0.1 (threshold evaluated in the Results section), the edge is discarded. However, we make a distinction when removing edges: if the edge is internal, all pixels of the edge are completely removed and, thus, the vertex-pixels at the end of that edge become edge-pixels; in contrast, if the edge is in contact with the non-ROI area, we only remove a few pixels in the middle of the edge in order to preserve the vertices. This is relevant because we use the vertices to determine the HEX. 

\textbf{(VI)} In the updated binary segmentation, every superpixel in contact with the border of the image is discarded. For the remaining superpixels, we checked the Body/ROI image to determine whether to keep or discard them. If using the body/blob image, a superpixel is included if the average intensity of their pixels is above 0.5. If using the ROI image, a superpixel is included if at least 85\% of its area is within the ROI.

\textbf{(VII)} In the final segmentation, the parameters are estimated. 

\begin{table*}[t]%[tbhp]
	\centering
	\caption{The average accuracy, DICE, and MHD in the 10-fold cross-validation scheme for the different networks (for CNN-Edge and CNN-Body), the number of parameters of each network, and their resulting MAE of the endothelial parameters.} 
	\begin{tabular}{lcccccccccc}
		& \multicolumn{3}{c}{CNN-Edge} & \multicolumn{3}{c}{CNN-Body} &  & ECD MAE & CV MAE & HEX MAE \\
		\cmidrule(lr){2-4}\cmidrule(lr){5-7}
		& Acc. [\%] & DICE [\%] & MHD [px] & Acc. [\%] & DICE [\%] & MHD [px] & Param. & [cells/mm$^2$] & [\%] & [\%]\\
		\midrule
		UNet                   & 96.37 & 86.16 & 0.129 & 94.25 & 92.02 & 0.563 & 21.59 M & 25.79 & 1.52 & 3.89 \\
		UNet+                  & 96.32 & 86.07 & 0.140 & 94.37 & 92.14 & 0.639 & 24.44 M & 24.83 & 1.61 & 3.79 \\
		UNet++                 & 96.30 & 86.00 & 0.141 & 94.40 & 92.19 & 0.566 & 24.44 M & 24.58 & 1.49 & 3.77 \\
		\midrule
		ResUNeXt               & 96.31 & 86.03 & 0.144 & 93.98 & 91.65 & 0.582 &  6.87 M & 25.84 & 1.69 & 3.79 \\
		ResUNeXt+              & 96.31 & 86.00 & 0.133 & 94.40 & 92.04 & 0.575 &  7.88 M & 26.66 & 1.53 & 3.75 \\
		ResUNeXt++             & 96.32 & 86.04 & 0.134 & 94.36 & 92.14 & 0.592 &  7.88 M & 26.49 & 1.67 & 3.75 \\
		\midrule
		DenseUNet              & 96.41 & \textbf{86.30} & 0.139 & 94.48 & 92.04 & 0.688 &  0.38 M & 23.81 & 1.30 & 3.36 \\
		DenseUNet+             & 96.40 & 86.26 & 0.139 & 94.54 & 92.10 & 0.729 &  0.35 M & 23.67 & 1.31 & 3.43 \\
		DenseUNet++            & \textbf{96.42} & \textbf{86.30} & 0.137 & 94.54 & 92.09 & 0.656 &  0.35 M & 24.77 & 1.43 & 3.43 \\
		DenseUNet Oktay        & 96.40 & 86.27 & 0.140 & 94.49 & 92.10 & 0.691 &  0.38 M & 23.50 & 1.37 & 3.26 \\
		\midrule
		DenseUNet fNLA Mul     & 96.40 & 86.29 & 0.125 & \textbf{94.69} & 92.36 & 0.578 &  0.43 M & \textbf{23.16} & 1.28 & 3.13 \\
		DenseUNet fNLA Conc    & 96.41 & \textbf{86.30} & 0.124 & 94.67 & \textbf{92.45} & \textbf{0.555} &  0.45 M & 24.05 & 1.37 & \textbf{3.12} \\
		DenseUNet fNLA Add     & 96.36 & 86.21 & 0.125 & 94.66 & 92.36 & 0.592 &  0.45 M & 24.72 & 1.32 & 3.15 \\
		\midrule
		DenseUNet++ fNLA Mul   & 96.38 & 86.25 & \textbf{0.119} & 94.60 & 92.15 & 0.669 &  0.43 M & 24.93 & \textbf{1.26} & 3.13 \\
		DenseUNet++ fNLA Conc  & 96.37 & 86.13 & 0.128 & 94.54 & 92.10 & 0.630 &  0.45 M & 24.43 & 1.40 & 3.27 \\
		DenseUNet++ fNLA Add   & 96.32 & 85.95 & 0.129 & 94.45 & 92.04 & 0.635 &  0.44 M & 26.18 & 1.43 & 3.30 \\
		\bottomrule
	\end{tabular}
	\label{table1}
\end{table*}

\subsection*{Implementation}

To evaluate the networks, a ten-fold cross-validation was performed (with all images from one eye in the same fold). All networks were implemented in Tensorflow 2.4.1 on a single NVIDIA V100 GPU with 32GB of memory. A training batch size of 15 images was employed, where six images were sampled from a specific guttae subgroup (two images from each complexity level). The dimensions of the DenseUNet were selected such that it could fit within the GPU memory; to build the multiscale DenseUNet+/++, we reduced the GR to 4 so that they could still fit in memory while having a similar number of parameters. In that respect, ResUNeXt+/++ and UNet+/++ had one less convolutional block in each resolution stage than the ResUNeXt and UNet (except the lowest stage). 

The network hyperparameters were categorical cross-entropy as loss function, Nadam optimizer \cite{Dozar2016}, initial learning rate ($lr$) of 0.001, no early stop, 200 epochs for CNN-Edge (learning decay of 0.99, considering that $lr_{epoch-i}=lr \cdot lr_{decay}^i$) and 100 epochs for CNN-Body, CNN-Blob, and CNN-ROI (learning decay of 0.97). For data augmentation, images were randomly flipped left-right and up-down, and elastic deformations were employed. The specular images were only normalized to have a range between 0--1 instead of 0--255. The networks were programmed in Python 3.7, and the parameter estimation and statistical analyses were done in Matlab 2020a (MathWorks, Natick, MA). 

The evaluation metrics for the CNN output images were accuracy, the S{\o}rensen-Dice coefficient (DICE), and modified Hausdorff distance (MHD) \cite{Dubuisson1994}. MHD is a good metric to evaluate the edge images because it measures how close the detected edges are from the manual annotations, whereas DICE can better evaluate the performance for the body images since it measures the overlap of the ROI. The evaluation metrics for the corneal parameters were mean absolute error (MAE) and mean absolute percentage error (MAPE). To assess the statistical significance between methods, we used the paired Wilcoxon test to compare the MAPE after assigning a 100\% error if no parameter estimate was produced. To assess the clinical statistical significance of our method against the gold standard, we used the Kruskal–Wallis H test. All tests used a statistical significance established at $\alpha$=0.05.

\section*{Results}

\subsection*{Comparison between types of CNNs}

\begin{figure*}%[t]
	\centering
	\includegraphics[width=1\linewidth]{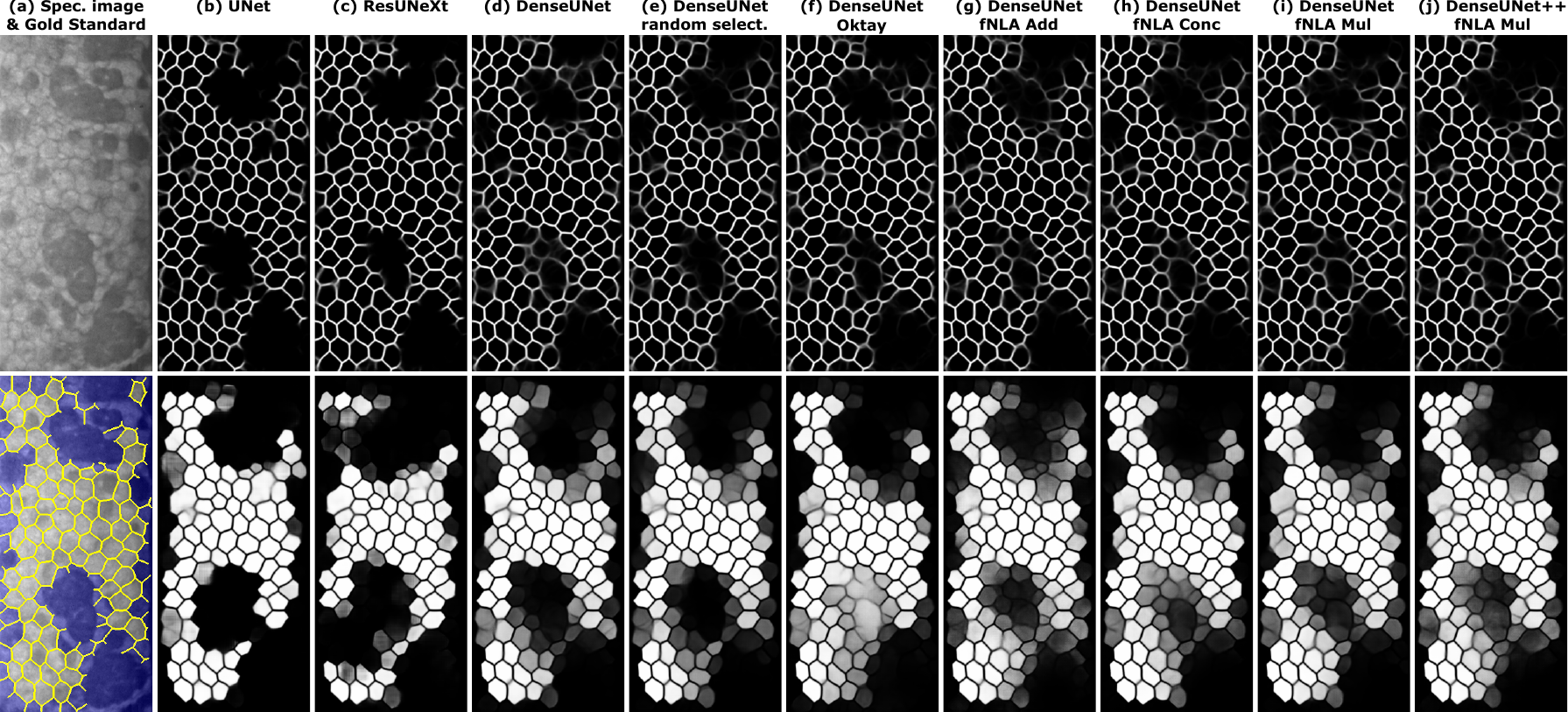}
	\caption{\textbf{(a)} Illustrative specular image with guttae (grade 5) and its manual annotation (edges in yellow; non-ROI area in blue). For the different networks (indicated on top), the inference of this image for the CNN-Edge \textbf{(top image)} and CNN-Body \textbf{(bottom image)}. Topcon's software did not detect a single cell in this image.}
	\label{fig04}
\end{figure*}

Quantitatively (Table~\ref{table1}), the basic UNet, ResUNeXt, and DenseUNet seemed to provide similar performance, with slightly better results for DenseUNet (more noticeable if we look at the MAE of the corneal parameters). However, the differences between those networks were discernible in the qualitative results (Fig.~\ref{fig04}): in the edge images, UNet and ResUNeXt provided rather binary results and were very conservative in the areas with large guttae (Fig.~\ref{fig04}-b, Fig.~\ref{fig04}-c, central area), whereas DenseUNet provided more probabilistic outputs and was able to extend the edge inference into those guttae areas, reaching beyond the manual annotations (Fig.~\ref{fig04}-d). In the case of the body images, DenseUNet showed a similar probabilistic behavior (Fig.~\ref{fig04}), whereas ResUNeXt had significant problems and the lowest performance (Table~\ref{table1}). Thus, DenseUNet was the network to further investigate.

One key point for the success of the edge inference within the guttae areas was the selective training procedure, where some of the images of the batch were sampled from specific guttae subsets. In fact, if the DenseUNets were trained without any stratified subsampling (random selection), such edge inferences did not appear (Fig.~\ref{fig04}-e). Another interesting observation was that the use of the multiscale networks (+ and ++) had no impact on the edge images but provided a clear improvement in the body images for all types of networks (Table~\ref{table1}), although this did not necessarily result in more accurate parameter estimations. % , knowing that 40\% of images have some guttae, thus it is still a well-balanced dataset

%Finally, it is worth mentioning that some hyperparameters were not pivotal but provided a slightly better performance: ELU activations yielded a minimally higher accuracy than ReLU in both edge and body images; BRN also provided slightly better results than the simpler batch normalization (BN) for body images but equal performance for the edge images; and the 20\% dropout was selected after a grid search, although any dropout value between 10--30\% would also have been acceptable (no use of dropout was markedly detrimental).

\subsection*{Selection of attention mechanism}

In 2018, Oktay et al. \cite{Oktay2018} proposed an attention mechanism in UNets, which modifies the output of a transition block by using the tensors from the lower resolution stage. In our experiments, we tested several variations of this mechanism, but none provided a clear improvement. The best design (Table~\ref{table1}, most like Oktay's proposal) gave sharper edge images in complicated areas but it occasionally provided poor results for the body images (Fig.~\ref{fig04}-f). 

In this work, we developed a different attention mechanism that would exploit the non-local dependencies between features of different resolution stages while simultaneously bringing back the attention information from the lowest stages back to the highest ones. Quantitatively, the improvement was subtle but perceptible in the basic DenseUNet: the MHD in the edge images was substantially smaller in all cases, and all metrics in the body images improved (Table~\ref{table1}). Qualitatively, they all provided sharper edge images (there were less double-edge artifacts within guttae areas, Fig.~\ref{fig04}), and the inference in the body images was more probabilistic (cells within guttae areas appeared with lower intensity instead of just pitch black). Thus, our attention mechanism seemed to correctly use the surrounding information to determine, in a probabilistic manner, whether a cell should be discarded. With regards to the type of aggregation, the concatenative and the multiplicative type provided similar results; in contrast, the additive type had the lowest performance in all the experiments. Therefore, we chose the multiplicative type because it uses fewer parameters and provides the overall smallest MAE in the corneal biomarkers (Table~\ref{table1}). On the other hand, the use of fNLA in the multiscale versions +/++ did not yield a clear benefit.

\subsection*{Postprocessing tuning}

The postprocessing had two hyperparameters to tune: a threshold to discard weak edges (`edge threshold'), and a threshold to discard superpixels from the final segmentation (`body threshold'). We performed a combined grid search and found that, for the edge threshold: (i) a value around 0.1 minimized the error for CV regardless of the body threshold; (ii) for HEX, the optimal value was 0.1 for a body threshold of 0.5 or lower, but it shifted towards 0.2 as we increased the body threshold towards 1; and (iii) for ECD, a higher threshold (0.2--0.3) was better but the differences were very small. Overall, the purpose of the edge threshold is to simply discard false edges created during the postprocessing (watershed). In that respect, a value of 0.1 seemed to be a reasonable choice.

\begin{figure*}[t]
	\centering
	\includegraphics[width=1\linewidth]{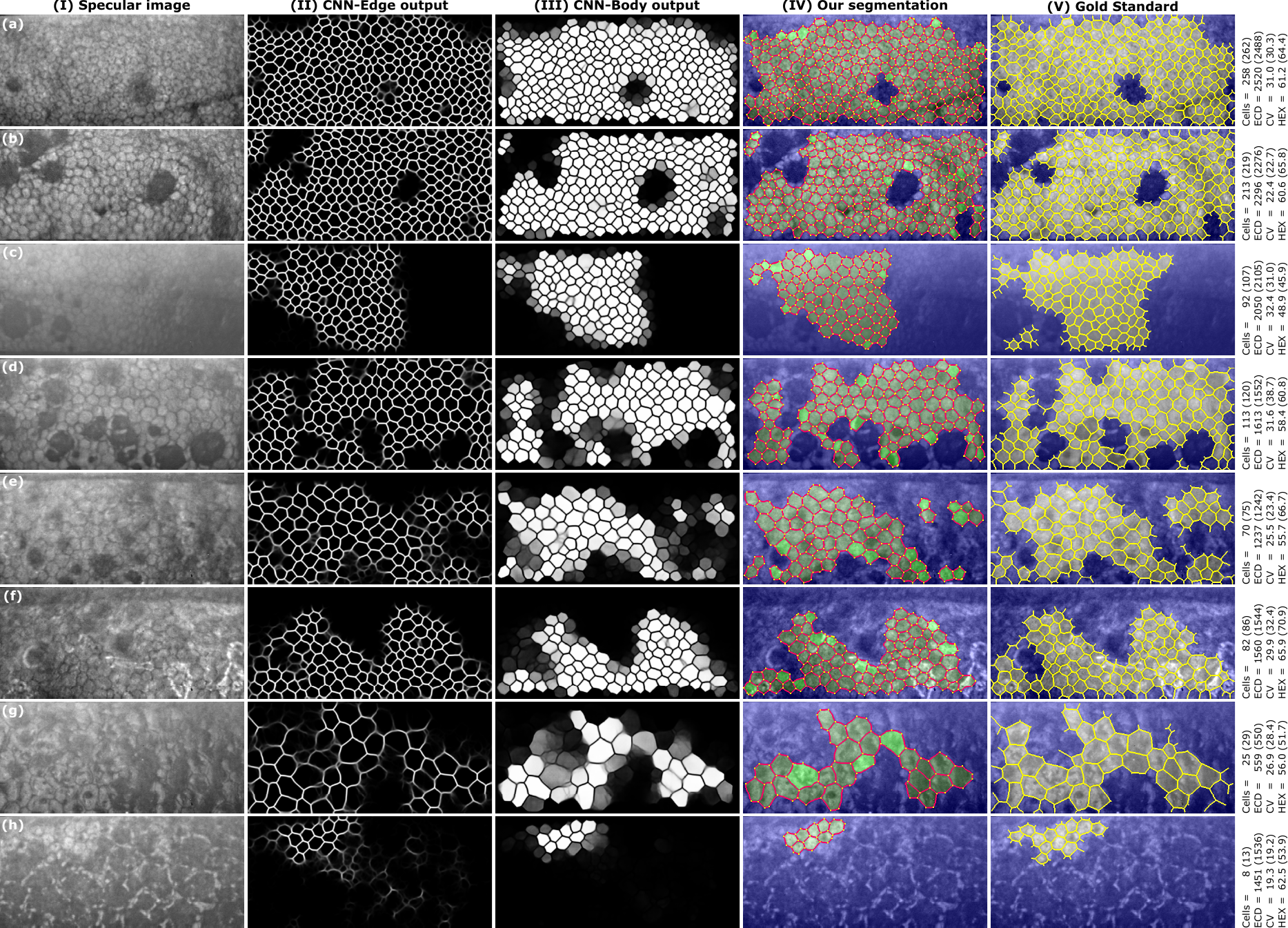}
	\caption{Eight images with guttae (two cases per grade): \textbf{(a, b)} grade 3, \textbf{(c, d)} grade 4, \textbf{(e, f)} grade 5, \textbf{(g, h)} grade 6. \textbf{(I)} The specular image; the output of the \textbf{(II)} CNN-Edge and \textbf{(III)} CNN-Body for the DenseUNet fNLA-Mul; \textbf{(IV)} the final segmentation (edges in red, vertices in yellow, non-ROI in blue, and the detected cells in two tonalities of green: brighter green for the cells whose body-intensity was between 0.50--0.75, and darker green for the cells whose body-intensity was between 0.75--1.00), and \textbf{(V)} the gold standard (the annotated edges in yellow; non-ROI in blue). On the right, the estimated parameters (gold standard values in parenthesis). Topcon's software detected 45, 34, 0, 38, 6, 0, 0, and 0 cells (a to h). }
	\label{fig05}
\end{figure*}

As for the body threshold, we observed that a value of 0.5 (which is the most intuitive choice) yielded the lowest errors in the three corneal parameters for the networks without attention, but it shifted towards 0.75 if we employed networks with fNLA, although the error differences were very small. However, visual inspection showed that no major mistakes were produced with a threshold of 0.5 (Fig.~\ref{fig05}, cells with brighter green). We believe that a lower threshold sometimes includes cells that were not annotated in the gold standard, therefore the differences do not indicate segmentation mistakes but mainly register the so-called `dissimilarity due to cell variability' (i.e. the estimated parameters can vary if a different set of cells are used to estimate them). %Thus, 0.5 was the selected value.

Finally, the postprocessing computed the hexagonality by using the cell vertices (vertex method) instead of counting the neighboring cells (neighbor method, which inevitably discards the cells in the periphery). For those peripheral cells, it was possible to detect their vertices if a portion of the peripheral edges was visible. This was true for most cases, although sometimes the vertex detection was faulty (Fig.~\ref{fig05}-IV-(e-f)). Nevertheless, the MAE in our HEX estimations was 3.14 [\%] with the vertex method and 4.13 [\%] with the neighbor method. Thus, the vertex method was better even when there were some mistakes in the detection of peripheral vertices.

%Finally, the postprocessing computed the hexagonality by using the cell vertices instead of counting the neighboring cells (the latter inevitably discards the cells in the periphery). For those peripheral cells, it was possible to detect their vertices as long as a small portion of the peripheral edges were detected. This was true for most cases, although sometimes the vertex detection was faulty (Fig.~\ref{fig05}-IV-(e-f)). Therefore, we evaluated whether it would still be better to use the old approach (neighbor method) to compute hexagonality instead of the new one (vertex method). First, we observed that, on the manual annotations, the mean absolute difference between the two types of estimations in HEX was 3.67 [\%] (the average number of cells per image in each type of estimation was 166 and 110 cells). Second, the MAE in our HEX estimations (compared against the vertex method in the gold standard) was 3.14 [\%] if we used the vertex method and 4.13 [\%] if we employed the neighbor method. Furthermore, if both sets used the neighbor method, the MAE in HEX was 5.07 [\%]. Therefore, the proposed new vertex method was significantly better even when there still were some mistakes in the detection of peripheral vertices.

\subsection*{Comparison between frameworks}

The performance of Topcon's algorithm was considerably inferior (Table~\ref{table2}), particularly for the images with guttae, where it failed to detect any cell in 30\% of the images (as in Fig.~\ref{fig05}-(c,f,g,h)) and it only detected one third of the cells (on average) in the remaining images.

\begin{table*}[t]%[tbhp]
	\centering
	\caption{The MAE of the endothelial parameters in the images with/without guttae, the percentage of images with estimates (success), and the average number of cells per image, for our methods and Topcon (the latter has two success percentages: one for ECV/CV and another for HEX). The manual annotations had an average of 148 and 166 cells for the images with/without guttae, respectively.}
	\begin{tabular}{lcccccccccc}
		&  \multicolumn{5}{c}{Images with guttae}  &  \multicolumn{5}{c}{Images without guttae}  \\
		\cmidrule(lr){2-6}\cmidrule(lr){7-11}
		& \% success & No. cells & ECD MAE & CV MAE & HEX MAE & \% success & No. cells & ECD MAE & CV MAE & HEX MAE\\
		\midrule
		Topcon            & 70\%/40\% &  53 & 166.81 & 4.54 & 9.31 & 87\%/60\% &  77 & 118.74 & 4.19 & 8.34 \\
		CNN-ROI method    & 99\%      & 157 & 42.47 & 2.15 & 6.59 & 100\%     & 175 & 21.59 & 1.13 & 3.11\\
		CNN-Blob method   & 96\%      & 143 & 36.93 & 2.04 & 4.87 & 100\%     & 164 & 17.68 & 0.97 & 2.54 \\
		CNN-Body method   & 97\%      & 143 & 34.66 & 1.98 & 4.58 & 100\%     & 164 & 17.64 & 0.98 & 2.50 \\
		\bottomrule
	\end{tabular}
	\label{table2}
\end{table*}

Regarding the previous approach CNN-ROI \cite{Vigueras2020}, the network had no problems to accurately infer the targets (Fig.~\ref{fig02}-j). However, the question was whether such targets are optimal to subsequently identify the well-detected cells. The error analysis indicated that the CNN-ROI method detected approximately 10--12 more cells per image, but there were a few segmentation mistakes among those cells, and the estimation errors were significantly worse than the CNN-Body approach (Table~\ref{table2}). The paired Wilcoxon test indicated a statistically significant difference between the estimates of approaches CNN-Body and CNN-ROI ($P<0.0001$, all biomarkers). % for the three biomarkers

As for the CNN-Blob, it detected the same number of cells as CNN-Body (Table~\ref{table2}) and the estimation errors were virtually the same (slightly worse for CNN-Blob in the images with guttae; Table~\ref{table2}). The paired Wilcoxon test indicated no statistically significant difference between the two estimates ($P=0.69$, $P=0.13$, and $P=0.69$, for the ECD, CV, and HEX, respectively). Nevertheless, it is easier for a human observer to interpret what cells are well detected in the CNN-Body output (Fig.~\ref{fig02}-h) than in the CNN-Blob output (Fig.~\ref{fig02}-i).

\subsection*{Statistical analysis}

The distributions of the estimated parameters from the CNN-Body method and the gold standard passed the Levene's test for homogeneity, but they did not pass the Shapiro-Wilk normality test. Thus, the Kruskal–Wallis H test was performed and it showed no statistically significant difference between the manual and our automatic assessments for ECD ($P=0.81$) and CV ($P=0.74$), but it did for HEX ($P<0.001$). To further assess the estimates, we performed a Bland-Altman analysis, and it showed that more than 95\% of the estimates were within the 95\% limit of agreement for all parameters: 96.5\% for ECD, 96.6\% for CV, and 96.7\% for HEX. % (using DenseUNets fNLA-Mul) 

\begin{figure*}[t]%[t]
	\centering
	\includegraphics[width=1\linewidth]{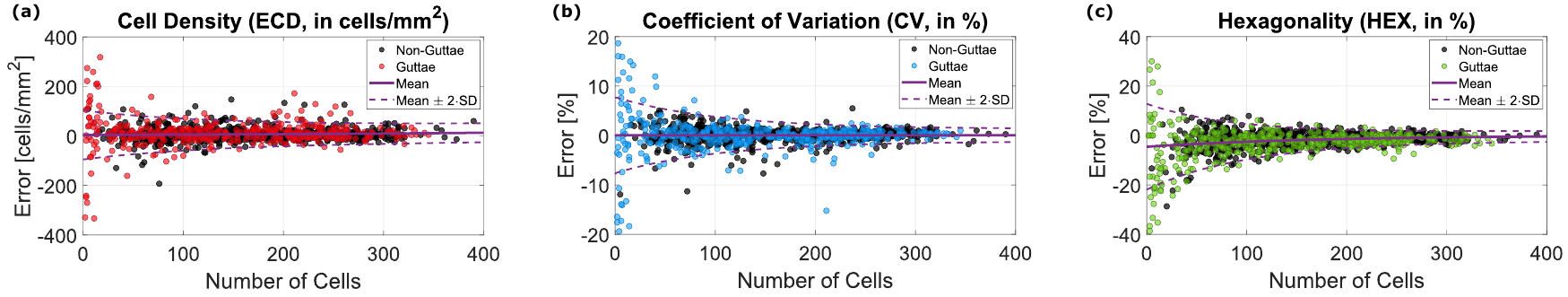}
	\caption{Error of our estimates of \textbf{(a)} ECD, \textbf{(b)} CV, and \textbf{(c)} HEX for the subsets of images without guttae (black dots, 702 images) and with guttae (colored circles, 483 images), displayed as a function of the number of cells. The y-axis indicates the error computed as the difference between the estimates and the gold standard. The mean (solid line) and two SDs (dashed lines) of the error function were modeled with exponentials.}
	\label{fig06}
\end{figure*}

We plotted the error as a function of the number of cells (Fig.~\ref{fig06}) and we fitted two exponentials to the mean and SD of the error using the least-squares method. The error estimates showed a normal distribution along the y-axis for the three parameters, which allowed us to assume that the area within two SDs covered approximately the 95\% of the error. This evaluation clearly showed that (i) the number of cells was the main variable to predict the reliability of the estimation, with a clear decrease in the error spread as more cells were detected; (ii) the unreliable cases were images with guttae and less than 25 cells; (iii) HEX required more cells to reduce the error spread, and (iv) there was a notable underestimation in HEX for the images with less than 100 cells. 

\subsection*{Clinical analysis}

The number of cells necessary to estimate ECD with high reliability is widely accepted as 75 cells \cite{Doughty2000}. At that point, our estimated ECD error was approximately 0 $\pm$ 35 [cells/mm$^2$] (mean $\pm$ SD). At 25 cells, it was 0 $\pm$ 45 [cells/mm$^2$]. This is expected to be much lower than the uncertainty generated by the "cell variability". Doughty et al. \cite{Doughty2000} evaluated that uncertainty (in manual annotations) and concluded that estimating ECD using 75 cells entails to assume an uncertainty (of 1 SD) of $\pm$2\% (or $\pm$70 cells/mm$^2$, which is twice than our method's error), whereas the uncertainty at 25 cells increases to $\pm$10\% (or $\pm$350 cells/mm$^2$, more than seven times larger than our method's error). Nevertheless, both elements (uncertainty and error) should be taken into consideration when evaluating the reliability of the estimations. Regarding CV and HEX, there are no studies in the literature about the effect of the cell variability to allow us to make a comparison. % This means that the uncertainty in ECD by the cell variability is twice than our method's error for images with 75 cells and more than seven times larger for images with 25 cells.

\begin{figure*}%[b]
	\centering
	\includegraphics[width=1\linewidth]{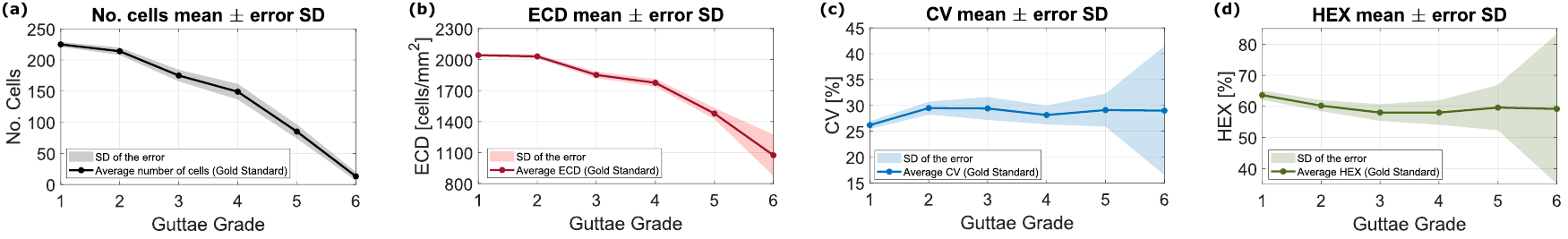}
	\caption{Average value of the biomarkers from the manual annotations --\textbf{(a)} number of cells, \textbf{(b)} ECD, \textbf{(c)} CV, and \textbf{(d)} HEX-- and the standard deviation of the estimated error for the different subsets of graded guttae images. The error SD allows to visualize how close the estimations are to the gold standard values.}
	\label{fig07}
\end{figure*}

We also observed that the number of visible cells and ECD decrease acutely as the amount of guttae increases (Fig.~\ref{fig07}), whereas neither CV nor HEX showed a significant change. While it is generally accepted that a CV of less than 30\% and a HEX of greater than 60\% is usually a sign of a healthy endothelium \cite{Vigueras2020}, this assumption seems to be untrue for the current dataset. Thus, our results seem to suggest that guttae does not have an impact on neither the cell size variation nor the hexagonality of the visible cells (those that are not actually affected by extensive guttae).

Overall, our proposed method has a very low error spread for the three biomarkers in images graded with mild or moderate guttae (Fig.~\ref{fig07}, up to grade 4). For the cases with severe guttae (grade 5--6), the main problem is the low number of detected cells (in grade 6, the average number of cells per image was 13 $\pm$ 10), which translates into a large estimation uncertainty, particularly for CV and HEX (Fig.~\ref{fig07}). As shown in Fig.~\ref{fig05}-(g,h), we did not observe major segmentation mistakes in those cases, but the low number of visible cells makes the biomarker estimation unreliable.

\section*{Discussion}

We have presented a fully automatic method for the estimation of the corneal endothelium parameters from specular microscopy images that contain guttae. This is the first time that such images have been solved properly in the literature. 

The main factor to achieve such accuracy was the way annotations were performed. We observed that, if the edges hidden by the guttae could be manually delineated, the networks could learn to replicate such pattern. Since endothelial cell edges usually appear as straight lines, the extension of partially occluded edges is sufficient to infer the hidden tessellation if the guttae are small enough. In fact, we observed that our model went beyond what our annotator could do. For example, the guttae indicated by the green arrow in Fig.~\ref{fig02} was too large to be certain about the hidden edges, yet the network provided a segmentation that seemed highly probable. We observed this behavior in many other cases, which made us conclude that artificial intelligence surpassed the human ability in this task. %Regarding the amount of data, this study contained the largest dataset of specular images to date. Considering that annotating a single specular image might take 30--60 minutes, it is understandable that most manuscripts in the literature only use a few hundreds of images at most. However, in this case, it was important to collect enough guttae cases while also providing sufficient healthy images to build a balance dataset.

%Our experiments clearly showed that DenseUNet was the most suitable type of network. We believe this is due to the peculiarities of the features in the images. Corneal endothelial images do not possess a large variety of unique features; instead, they depict many variations of the same type of feature (an edge). Since the strength of DenseUNets lies in the reuse of previous feature maps within a single dense block, it is a good match for this problem. 

Our proposed attention mechanism (fNLA) moderately improved the performance in both types of CNNs (CNN-Edge and CNN-Body). Since the basic DenseUNet already achieved an excellent performance, the improvement provided by our attention block was modest. Nevertheless, we believe that the feedback-attention path depicted in our proposed network could yield a good performance in other types of segmentation problems, particularly the ones that infer areas instead of edges. %  whereas a well-known attention UNet in the literature \cite{Oktay2018} could not provide any benefit and was actually detrimental sometimes

One interesting behavior of this framework is that, while CNN-Edge provides good inference beyond the manual annotations, CNN-Body is rather conservative. This is because the targets of CNN-Body were based on the original annotations and, thus, any inference by CNN-Edge that surpasses the manual annotations are not considered in the target of CNN-Body. While this might seem a negative quality, the results show that our framework detects practically the same number of cells as the manual annotations and barely any segmentation mistake is observed even in the most difficult cases (Fig.~\ref{fig05}). Therefore, we believe this approach is preferred to more daring ones, like CNN-ROI (Table~\ref{table2}). 

%Another peculiarity is that CNN-Body identifies the well-detected cells independently of the surrounding cells, depicting uniform intensity within the area of each cell. This suggests that the network clearly understands where the limits of each cell are. However, sometimes unusual results occur; for instance in Fig.~\ref{fig02}-h (inner cell in black) a single cell is not classified as trustworthy even when it is completely surrounded by well-detected cells. In that case, the network could probably not ascertain whether this was one or two cells and, thus, decided to discard it to avoid any possible mistake. In that respect, CNN-Body showed a conservative behavior.

%On a different aspect, CNN-Edge was exceptionally efficient to detect edges in very complicated images while not providing noisy results (Fig.~\ref{fig05}-II). In our previous work \cite{Vigueras2020}, we could observe that, for images with very large guttae (similar to the ones in Fig.~\ref{fig05}-(g,h)), the network often provided blurry, false edges where guttae were present, which subsequently distorted the computations in the postprocessing, thereby yielding wrong segmentations. In contrast, the edge detection shows now a very low fall-out.

Overall, our estimates agreed very well with the gold standard and they were significantly better than the ones provided by the instrument's software, demonstrating the ability of this artificial intelligence framework to accurately estimate the endothelial parameters from images with guttae.

\section*{Code availability}
Code for the networks and their weights are available at: \url{https://github.com/jpviguerasguillen/feedback-non-local-attention-fNLA}. %Due to regulatory restrictions, we cannot make the data open-access at this moment.

\acknow{We thank Angela Engel, Caroline Jordaan, and Annemiek Krijnen for their contribution in acquiring the images.}

\showacknow % Display the acknowledgements section

\section*{References}

%\bibliography{fuchs_paper}

\begin{thebibliography}{10}

\bibitem{Elhalis2010}
H Elhalis, B Azizi, UV Jurkunas, Fuchs endothelial corneal dystrophy.
\newblock {\em\protect\JournalTitle{The Ocular Surface}} \textbf{8}, 173–184
  (2010).

\bibitem{McLaren2014}
JW McLaren, LA Bachman, KM Kane, SV Patel, Objective assessment of the corneal
  endothelium in {F}uchs' endothelial dystrophy.
\newblock {\em\protect\JournalTitle{Investigative Ophthalmology \& Visual
  Science}} \textbf{55}, 1184--1190 (2014).

\bibitem{Gain2016}
P Gain, et~al., Global survey of corneal transplantation and eye banking.
\newblock {\em\protect\JournalTitle{JAMA Ophthalmology}} \textbf{134}, 167--176
  (2016).

\bibitem{Adamis1993}
AP Adamis, V Filatov, BJ Tripathi, RC Tripathi, Fuchs' endothelial dystrophy of
  the cornea.
\newblock {\em\protect\JournalTitle{Survey of Ophthalmology}} \textbf{38},
  149--168 (1993).

\bibitem{Moshirfar2021}
M Moshirfar, AN Somani, U Vaidyanathan, BC Patel, Fuchs endothelial dystrophy
  (In: StatPearls Publishing) (2021) Available from:
  https://www.ncbi.nlm.nih.gov/books/NBK545248/.

\bibitem{Foster2004}
CS Foster, DT Azar, CH Dohlman, {\em Smolin and {T}hoft’s the cornea:
  scientific foundations \& clinical practice}.
\newblock (Lippincott Williams \& Wilkins, Philadelphia, PA), Fourth edition,
  (2004) 46--48.

\bibitem{McCarey2008}
BE McCarey, HF Edelhauser, MJ Lynn, Review of corneal endothelial specular
  microscopy for {FDA} clinical trials of refractive procedures, surgical
  devices, and new intraocular drugs and solutions.
\newblock {\em\protect\JournalTitle{Cornea}} \textbf{27}, 1--16 (2008).

\bibitem{Huang2018}
J Huang, et~al., Comparison of noncontact specular and confocal microscopy for
  evaluation of corneal endothelium.
\newblock {\em\protect\JournalTitle{Eye \& Contact Lens}} \textbf{44},
  S144--S150 (2018).

\bibitem{Price2013}
MO Price, KM Fairchild, FW Price~Jr, Comparison of manual and automated
  endothelial cell density analysis in normal eyes and {DSEK} eyes.
\newblock {\em\protect\JournalTitle{Cornea}} \textbf{32}, 567--873 (2013).

\bibitem{Luft2015}
N Luft, N Hirnschall, S Schuschitz, P Draschl, O Findl, Comparison of 4
  specular microscopes in healthy eyes and eyes with cornea guttata or corneal
  grafts.
\newblock {\em\protect\JournalTitle{Cornea}} \textbf{34}, 381--386 (2015).

\bibitem{Gasser2015}
L Gasser, T Reinhard, D B\"{o}hringer, Comparison of corneal endothelial cell
  measurements by two non-contact specular microscopes.
\newblock {\em\protect\JournalTitle{BMC Ophthalmology}} \textbf{15}, 87 (2015).

\bibitem{Kitzmann2004}
AS Kitzmann, et~al., Comparison of corneal endothelial cell images using a
  noncontact specular microscope and the confoscan 3 confocal microscope.
\newblock {\em\protect\JournalTitle{Investigative Ophthalmology \& Visual
  Science}} \textbf{45}, 155 (2004).

\bibitem{Piorkowski2015}
A Pi\'{o}rkowski, J Gronkowska-Serafin, Towards precise segmentation of corneal
  endothelial cells in {\em International Conference on Bioinformatics and
  Biomedical Engineering (IWBBIO 2015).}
\newblock (Granada, Spain), Vol.{} LNCS 9043, pp. 240--249 (2015).

\bibitem{Selig2015}
B Selig, KA Vermeer, B Rieger, T Hillenaar, CL Luengo~Hendriks, Fully automatic
  evaluation of the corneal endothelium from in vivo confocal microscopy.
\newblock {\em\protect\JournalTitle{BMC Medical Imaging}} \textbf{15}, 13
  (2015).

\bibitem{Scarpa2016}
F Scarpa, A Ruggeri, Development of a reliable automated algorithm for the
  morphometric analysis of human corneal endothelium.
\newblock {\em\protect\JournalTitle{Cornea}} \textbf{35}, 1222--1228 (2016).

\bibitem{Al-Fahdawi2018}
S Al-Fahdawi, et~al., A fully automated cell segmentation and morphometric
  parameter system for quantifying corneal endothelial cell morphology.
\newblock {\em\protect\JournalTitle{Computer Methods and Programs in
  Biomedicine}} \textbf{160}, 11--23 (2018).

\bibitem{Vigueras2018}
JP Vigueras-Guill\'{e}n, et~al., Corneal endothelial cell segmentation by
  classifier-driven merging of oversegmented images.
\newblock {\em\protect\JournalTitle{IEEE Transactions on Medical Imaging}}
  \textbf{37}, 2278--2289 (2018).

\bibitem{Fabijanska2018}
A Fabija\'{n}ska, Segmentation of corneal endothelium images using a
  {U-Net-based} convolutional neural network.
\newblock {\em\protect\JournalTitle{Artificial Intelligence in Medicine}}
  \textbf{88}, 1--13 (2018).

\bibitem{Nurzynska2018}
K Nurzynska, Deep learning as a tool for automatic segmentation of corneal
  endothelium images.
\newblock {\em\protect\JournalTitle{Symmetry}} \textbf{10}, 60 (2018).

\bibitem{Daniel2019}
MC Daniel, et~al., Automated segmentation of the corneal endothelium in a large
  set of 'real-world' specular microscopy images using the {U-net}
  architecture.
\newblock {\em\protect\JournalTitle{Nature Scientific Reports}} \textbf{9},
  4752 (2019).

\bibitem{Fabijanska2019}
A Fabija\'{n}ska, Automatic segmentation of corneal endothelial cells from
  microscopy images.
\newblock {\em\protect\JournalTitle{Biomedical Signal Processing and Control}}
  \textbf{47}, 145--148 (2019).

\bibitem{Kolluru2019}
C Kolluru, et~al., Machine learning for segmenting cells in corneal endothelium
  images in {\em Proceedings of SPIE 2019}.
\newblock (San Diego, CA, USA), Vol.{} 10950, p. 109504G (2019).

\bibitem{Vigueras2019a}
JP Vigueras-Guill\'{e}n, et~al., Fully convolutional architecture vs
  sliding-window {CNN} for corneal endothelium cell segmentation.
\newblock {\em\protect\JournalTitle{BMC Biomedical Engineering}} \textbf{1}, 4
  (2019).

\bibitem{Vigueras2019b}
JP Vigueras-Guill\'{e}n, HG Lemij, J van Rooij, KA Vermeer, LJ van Vliet,
  Automatic detection of the region of interest in corneal endothelium images
  using dense convolutional neural networks in {\em Proceedings of SPIE,
  Medical Imaging: Image Processing}.
\newblock (San Diego, CA, USA), Vol.{} 10949, p. 1094931 (2019).

\bibitem{Vigueras2019c}
JP Vigueras-Guill\'{e}n, J van Rooij, HG Lemij, KA Vermeer, LJ van Vliet,
  Convolutional neural network-based regression for biomarker estimation in
  corneal endothelium microscopy images in {\em 41st Annual International
  Conference of the IEEE Engineering in Medicine and Biology Society (EMBC)}.
\newblock (Berlin, Germany), pp. 876--881 (2019).

\bibitem{Joseph2020}
N Joseph, et~al., Quantitative and qualitative evaluation of deep learning
  automatic segmentations of corneal endothelial cell images of reduced image
  quality obtained following cornea transplant.
\newblock {\em\protect\JournalTitle{Journal of Medical Imaging}} \textbf{7},
  014503 (2020).

\bibitem{Sierra2020}
JS Sierra, et~al., Automated corneal endothelium image segmentation in the
  presence of cornea guttata via convolutional neural networks. in {\em
  Proceedings of the SPIE, Applications of Machine Learning}.
\newblock (Online), Vol.{} 11511, p. 115110H (2020).

\bibitem{Vigueras2020}
JP Vigueras-Guill\'{e}n, et~al., Deep learning for assessing the corneal
  endothelium from specular microscopy images up to 1 year after
  {ultrathin-DSAEK} surgery.
\newblock {\em\protect\JournalTitle{Translational Vision Science \&
  Technology}} \textbf{9}, 49 (2020).

\bibitem{Karmakar2021}
R Karmakar, S Nooshabadi, A Eghrari, An automatic approach for cell detection
  and segmentation of corneal endothelium in specular microscope.
\newblock {\em\protect\JournalTitle{Graefes Arch Clin Exp Ophthalmol}}
  \textbf{Nov 6} (2021).

\bibitem{Kucharski2021}
A Kucharski, A Fabija\'{n}ska, {CNN-watershed}: a watershed transform with
  predicted markers for corneal endothelium image segmentation.
\newblock {\em\protect\JournalTitle{Biomedical Signal Processing and Control}}
  \textbf{68}, 102805 (2021).

\bibitem{Shilpashree2021}
PS Shilpashree, KV Suresh, RR Sudhir, SP Srinivas, Automated image segmentation
  of the corneal endothelium in patients with {Fuchs dystrophy}.
\newblock {\em\protect\JournalTitle{Translational Vision Science \&
  Technology}} \textbf{10}, 27 (2021).

\bibitem{Herrera2021}
R Herrera-Pereda, A Taboada~Crispi, D Babin, W Philips, M Holsbach~Costa, A
  review on digital image processing techniques for in-vivo confocal images of
  the cornea.
\newblock {\em\protect\JournalTitle{Medical Image Analysis}} \textbf{73},
  102188 (2021).

\bibitem{Ronneberger2015}
O Ronneberger, P Fischer, T Brox, {U-Net}: convolutional networks for
  biomedical image segmentation in {\em Medical Image Computing and
  Computer-Assisted Intervention (MICCAI)}.
\newblock (Munich, Germany), Vol.{} LNCS 9351, pp. 234--241 (2015).

\bibitem{Xie2017}
S Xie, R Girshick, P Doll\'{a}r, Z Tu, K He, Aggregated residual
  transformations for deep neural betworks in {\em 2017 IEEE Conference on
  Computer Vision and Pattern Recognition (CVPR)}.
\newblock (Honolulu, USA), pp. 5987--5995 (2017).

\bibitem{Jegou2017}
S J\'{e}gou, M Drozdzal, D Vazquez, A Romero, Y Bengio, The one hundred layers
  tiramisu: fully convolutional {DenseNets} for semantic segmentation in {\em
  2017 IEEE Conference on Computer Vision and Pattern Recognition (CVPR)
  Workshops}.
\newblock (Honolulu, USA), pp. 1175--1183 (2017).

\bibitem{Zhou2020}
Z Zhou, MMR Siddiquee, N Tajbakhsh, J Liang, {UNet++}: redesigning skip
  connections to exploit multiscale features in image segmentation.
\newblock {\em\protect\JournalTitle{IEEE Transactions on Medical Imaging}}
  \textbf{39}, 1856--1867 (2020).

\bibitem{Oktay2018}
O Oktay, et~al., Attention {U-Net}: learning where to look for the pancreas in
  {\em 1st conference on Medical Imaging with Deep Learning (MIDL)}.
\newblock (Amsterdam, The Netherlands), (2018).

\bibitem{Vigueras2018b}
JP Vigueras-Guill\'{e}n, et~al., Improved accuracy and robustness of a corneal
  endothelial cell segmentation method based on merging superpixels in {\em
  15th International Conference on Image Analysis and Recognition (ICIAR
  2018)}.
\newblock (Póvoa de Varzim, Portugal.), Vol.{} LNCS 10882, pp. 631--638
  (2018).

\bibitem{Wang2018}
X Wang, R Girshick, A Gupta, K He, Non-local neural networks in {\em 2018
  IEEE/CVF Conference on Computer Vision and Pattern Recognition (CVPR)}.
\newblock (Salt Lake City, USA), pp. 7794--7803 (2018).

\bibitem{Ioffe2016}
S Ioffe, Batch renormalization: towards reducing minibatch dependence in
  batch-normalized models in {\em 31st Conference on Neural Information
  Processing Systems (NIPS)}.
\newblock (Long Beach, USA), (2016).

\bibitem{Clevert2016}
DA Clevert, T Unterthiner, S Hochreiter, Fast and accurate deep network
  learning by exponential linear units {(ELUs)} in {\em International
  Conference on Learning Representations (ICLR)}.
\newblock (San Juan, Puerto Rico), (2016).

\bibitem{Srivastava2014}
N Srivastava, G Hinton, A Krizhevsky, I Sutskever, R Salakhutdinov, Dropout: a
  simple way to prevent neural networks from overfitting.
\newblock {\em\protect\JournalTitle{Journal of Machine Learning Research}}
  \textbf{15}, 1929--1958 (2014).

\bibitem{Vaswani2017}
A Vaswani, et~al., Attention is all you need in {\em 31st Conference on Neural
  Information Processing Systems (NIPS)}.
\newblock (Long Beach, USA), (2017).

\bibitem{Srinivas2021}
A Srinivas, et~al., Bottleneck transformers for visual recognition in {\em 2021
  IEEE/CVF Conference on Computer Vision and Pattern Recognition (CVPR)}.
\newblock (Nashville, TN, USA), pp. 16514--16524 (2021).

\bibitem{Beucher1993}
S Beucher, F Meyer, {\em The morphological approach to segmentation: the
  watershed transformation. Mathematical morphology in image processing,
  chapter 12.}
\newblock (Marcel Dekker), (1993) pp. 433--481.

\bibitem{Dozar2016}
T Dozat, Incorporating {Nesterov} momentum into {Adam} in {\em International
  Conference on Learning Representations (ICLR) Workshop}.
\newblock (San Juan, Puerto Rico), Vol.{}~1, pp. 2013--2016 (2016).

\bibitem{Dubuisson1994}
MP Dubuisson, AK Jain, A modified {Hausdorff} distance for object matching in
  {\em Proceedings of 12th International Conference on Pattern Recognition}.
\newblock (Jerusalem, Israel), pp. 566--568 (1994).

\bibitem{Doughty2000}
MJ Doughty, A M{\"u}ller, ML Zaman, Assessment of the reliability of human
  corneal endothelial cell-density estimates using a noncontact specular
  microscope.
\newblock {\em\protect\JournalTitle{Cornea}} \textbf{19}, 148--158 (2000).

\end{thebibliography}

\end{document}